\documentclass[12pt,a4paper]{JHEP3}

\usepackage{amscd,amsmath,amssymb,amsfonts,xspace,mathtools}
\usepackage{xcolor}
\usepackage{tikz-cd}
\usepackage{url}
\let\normalcolor\relax      
\usepackage{latexsym}  
\usepackage{graphicx}

\usepackage{graphicx, subfig, float}

\usepackage{epsfig,amsfonts,amssymb}
\usepackage{framed}

\newcommand{\refb}[1]{(\ref{#1})}
\newcommand{\Tr}{\mathop{\mathrm{Tr}}}

\newcommand{\nn}{\ensuremath{\nonumber{}}}

\title{An  extremal black hole with a unique ground state}

\author{Swapnamay Mondal$^{1,2}$
\\
{\it $~^1$ Department of Physics, Institute of Science, Banaras Hindu University, Varanasi, 221005, India \\
$~^2$ Dublin Institute for Advanced Studies, 10 Burlington Road, Dublin, Ireland} 
\vspace*{2mm}\\
{\tt swapno@maths.tcd.ie }
\vspace*{-3mm}
}

\abstract{
Recent computations in gravity suggest that non-supersymmetric extremal black holes lack any sizeable ground state degeneracy. We confirm this for D-brane description of non-supersymmetric 4-charge extremal black holes in N=8 string theory.  The microscopic description comprises four stacks of D-branes wrapping various cycles of the internal six-torus and intersecting at a point. The orientations of the stacks are such that supersymmetry is broken completely. We construct the low energy worldline Lagrangian for the brane system, which is seen to have 32 Goldstinos and 28 Goldstones. The Hamiltonian has a unique ground state, which carries a non-zero energy implying the absence of any truly extremal state.\\

\today}

\begin{document}
 
\maketitle

%
%
%
%
%
%
%
%
%
%
\section{Introduction}
As per semi-classical gravity, extremal black holes carry vanishing temperature. Then standard statistical mechanics entails that the ground state degeneracy of extremal black holes should be exponential of its entropy. This is a huge degeneracy and begs the question: what stops this degeneracy from getting lifted? For supersymmetric black holes, the answer would be supersymmetry. But for non-supersymmetric black holes, the answer is less than clear.  

This issue has been debated for quite sometime. E.g. Page has suggested there are no true degeneracies for non-supersymmetric  extremal black hole \cite{Page:2000dk}. In fact the semi-classical analysis also indicates the presence of a gap and it has been suggested that Hawking's analysis itself may not be trustable below this gap \cite{Preskill:1991tb, Maldacena:1998uz}.  
Recently, for near-extremal Reissner-Nordstr\"{o}m black holes, it has been shown that for low temperatures black hole entropy $S \sim \log{} T$, where $T$ is the temperature and the proportionality constant is positive \cite{Iliesiu:2020qvm}. This relation implies a vanishing density of states near the ground state. 
These results are hard to  reconcile with a sizeable ground state degeneracy. 


On the other hand, $S \sim \log{} T$ implies negative infinite entropy at zero temperature, which is hard to reconcile with a discreet spectrum, typical of any quantum system with finitely many degrees of freedom\footnote{The Schwarzian theory \cite{Kit1, Kit2, Maldacena:2016hyu, Maldacena:2016upp, Engelsoy:2016xyb, Stanford:2017thb, Mertens:2017mtv, Lam:2018pvp, Kitaev:2018wpr, Yang:2018gdb, Saad:2019lba, Iliesiu:2019xuh,Choi:2021nnq, Choi:2023syx}, which this relation emanates from, has been reproduced in CFT descriptions \cite{Ghosh:2019rcj}, but the CFT descriptions describe an effective string capturing an infinite number of black holes at one go, hence has infinitely many degrees of freedom. }. There have been indications that $S \sim \log{} T$ behaviour is trustable only in a ``low but not too low" temperature regime \cite{Mondal:2023bcx}. Overall the true nature of the ground state and low lying spectra remains elusive. In this article, we decisively settle the issue by looking at a microscopic description of a non-supersymmetric extremal black hole. 
  
Most well known microscopic descriptions are two dimensional conformal field theories (CFTs) capturing the dynamics of an effective string wrapping a circle \cite{Strominger:1996sh, Maldacena:1997de}. Black holes, including non-supersymmetric ones appear as highly degenerate states in a CFT. Entropies of a great many non-supersymmetric black holes have been accounted for this way \cite{Breckenridge:1996sn, Maldacena:1996iz, Horowitz:1996ay, Sfetsos:1997xs, Maldacena:1997nx, Maldacena:1996ya, Horowitz:1996ac, Dabholkar:1997rk, Danielsson:2001xe, Cvetic:1996dt, Cvetic:1998vb, Cvetic:1998hg, Das:1997kt, Zhou:1996nz}. 
However this degeneracy is likely a consequence of conformal symmetry, rather than extremality. E.g. even non-extremal states have exponential degeneracies \cite{Horowitz:1996ay, Horowitz:1996ac,Danielsson:2001xe}. Furthermore the relevant CFTs are free which precludes any important effect interactions could have had\footnote{Wee thank Roberto Emparan for drawing our attention to this issue.}. 
If one could consider the full fledged sigma model describing the effective string (e.g. \cite{Minasian:1999qn} for the CFT in \cite{Maldacena:1997de}), perhaps the true origin of this degeneracy could be deciphered.

We take a different route. We consider a microscopic description where the low energy physics of the brane system is that of an effective particle, captured by a quantum mechanics. Interactions are captured in the potential. Unlike in a CFT, large degeneracy is atypical of quantum mechanics and hence such microscopic descriptions are perfect for settling this issue. Such a description was considered in \cite{Emparan:2006it}, but the state counting amounted to counting brane intersections  and hence was oblivious to the dynamics. Instead we exploit the  $D2  \textendash D2  \textendash D2  \textendash D6$ black hole, for which the explicit microscopic Hamiltonian was constructed in \cite{Chowdhury:2014yca, Chowdhury:2015gbk}. Our strategy would be to flip the orientation of one of the constituent D-brane stacks, say the $D6$ stack, so as to break supersymmetry completely \cite{Ortin:1996bz}. A closely related system was studied in \cite{Gimon:2007mh}, which realized non-BPS extremal black holes as threshold bound states of BPS branes. Also see \cite{Gimon:2009gk}. 
But no microscopic Hamiltonian was constructed in these works. In the present case, despite the lack of supersymmetry, we are able to determine the Hamiltonian to a significant extent, owing to supersymmetric nature of individual stacks. The resultant theory hosts 32 Goldstinos due to 32 broken supercharges of type $IIA$ string theory. The theory makes the absence of any ground degeneracy explicit. Also, the non-zero energy of the ground state implies that quantum mechanically there are no truly extremal non-supersymmetric states. 

The paper is organized as follows. In Section \ref{d2-d2-d2-d6} we quickly review the supersymmetric $D2  \textendash D2  \textendash D2  \textendash D6$ system. In section \ref{snsusy} We move on to analyze the $D2  \textendash D2  \textendash D2  \textendash \overline{D6}$ system. In section \ref{sdisc} we discuss implications of our results and future directions.


\section{The 4-charge BPS black hole} \label{d2-d2-d2-d6}
To begin with, we quickly gather the essentials of the $D2 \textendash D2  \textendash D2  \textendash D6$ system in type IIA string theory on $\mathbb{R}^{3,1} \times T^6$. The three D2 branes wrap three disjoint two-cycles of the $T^6$, say $x^4 \textendash x^5, \,  x^6 \textendash x^7$ and $x^8 \textendash x^9$. The D6 branes wrap the entire $T^6$. The branes intersect at a point, hence the low energy dynamics is that of an effective particle. Number of branes in various stacks amounts to charges carried by the black hole. The number of degrees of freedom of this effective particle grows quadratically with charges. For large charges the entropy of the corresponding black hole is $2\pi \sqrt{N_1 N_2 N_3 N_4}$, where $N_1, N_2, N_3, N_4$ respectively are the number of branes in four stacks \cite{Shih:2005qf}.

In order to write down the worldline theory of the effective particle, another crucial piece of information is the amount of supersymmetry preserved by the system. The 10 dimensional type IIA string theory has the supercharges 
\begin{align}
Q &= \overline{\epsilon}_L Q_L + \overline{\epsilon}_R Q_R
\end{align}
where $Q_{L.R}$ are the left/right moving supercharges and $\epsilon_{L,R}$ are ten dimensional Majorana-Weyl spinors with opposite chirality. Four stack of D-branes impose the following four conditions on  $\epsilon_{L,R}$:
\begin{align}
\nn
\epsilon_L &= s_{45} \Gamma^0 \Gamma^4 \Gamma^5 \epsilon_R \, , \\
\nn
\epsilon_L &= s_{67} \Gamma^0 \Gamma^6 \Gamma^7 \epsilon_R \, , \\
\nn
\epsilon_L &= s_{89} \Gamma^0 \Gamma^8 \Gamma^9 \epsilon_R \, , \\
\epsilon_L &= s_{456789} \Gamma^0 \Gamma^4 \Gamma^5 \Gamma^6 \Gamma^7 \Gamma^8 \Gamma^9 \epsilon_R \, , 
\end{align}
where $s$-s are signs and $\Gamma$-s are ten dimensional Gamma matrices. Simultaneous satisfaction of these conditions require
\begin{align}
s_{45} s_{67} s_{89} &= - s_{456789} \, , \label{smatch}
\end{align}
in which cases, 4 supercharges are preserved. In remaining cases, no supercharges exist. Note that the individual signs are not very consequential for the worldline theory, only the relative signs are.
To understand how the supersymmetry plays out at the level of the worldline Lagrangian, it is instructive to add stacks one by one. Since the theory is finally going to be reduced to 0+1 dimension, we do so to start with.

\paragraph{A single stack of D-branes:}
A single stack preserves 16 supercharges. The dynamics of zero-modes is same as the dynamics of D0 branes, captured by the BFSS model  \cite{Banks:1996vh}. Alternatively one can think of the dynamics as zero mode dynamics of a stack of D3 branes, captured by $\mathcal{N}=4$ super Yang Mills theory \cite{Witten:1995im}. We find it more convenient to use the language of four dimensional supersymmetry, with the implicit assumption that the fields depend only on time.

A $\mathcal{N}=4$ multiplet consists of one $\mathcal{N}=1$ vector multiplet $V=(A^\mu, \lambda)$ and three $\mathcal{N}=1$ chiral multiplets $\Phi_i = (\phi_i, \psi_i), \, i=1,2,3$. Here we have expressed the on-shell field contents as (boson, fermion). The complex bosons $\phi_1, \phi_2, \phi_3$ respectively describe the complex coordinate/Wilson line of/along the $45, 67, 89$ two-cycles\footnote{E.g. for the D2 brane wrapping the 45 directions, $\phi_1$ will be the Wilson line along 45, whereas $\phi_2, \phi_3$ will be the transverse coordinates along $67, 89$. For the D6 branes all three scalars describe Wilson lines.}. The periodicity of the torus coordinates/Wilson lines will be ignored, as we assume the branes never fluctuate enough to see the compact nature of the two tori. In fact the six real scalars are related by $SO(6)$ rotations. 
The spatial components of the vector field $A^\mu$ describe the transverse coordinates of the branes along three flat directions. 

The $\mathcal{N}=4$ superalgebra has a $SU(4)_R$ R-symmetry. This is reflected in the rotation of  four fermions, coming from four $\mathcal{N}=1$ multiplets as well as in the $SO(6) \sim SU(4)$ rotation of scalars. 
The overall $U(1)$ parts of the four fermions, coming from four $\mathcal{N}=1$ multiplets, lead to 16 kappa symmetries signifying 16 broken supersymmetries.

The superpotential is 
\begin{align}
W_{\mathcal{N}=4} &= \sqrt{2} \Tr{} \Phi_1 [ \Phi_2 , \Phi_3 ] \, .  
\end{align}
Note the   $\mathcal{N}=4$ super Yang Mills is oblivious to the orientation of the brane, which becomes evident in its  coupling to the Ramond Ramond forms. But this would not concern us, as only the relative orientation of different  stacks matter for our purpose.  

For rest of the paper, we will take each stack to contain a single brane, rendering the overall gauge group to be $U(1) \times U(1) \times U(1) \times U(1)$. Thus all fields will be numbers.

%

\paragraph{Two intersecting stacks of D-branes:}
Another stack of D-branes breaks half of the supersymmetry, preserved by the first brane. Thus a total 8 supercharges are preserved, amounting to $\mathcal{N}=2$ supersymmetry in four dimensions. The $\mathcal{N}=4$ multiplet (say of the first brane) can be decomposed in one $\mathcal{N}=2$ vector and another $\mathcal{N}=2$ hypermultiplet. An $\mathcal{N}=2$ vector multiplet contains the $\mathcal{N}=1$ vector $V$ and another $\mathcal{N}=1$ chiral, which could be any one of $\Phi_1, \Phi_2, \Phi_3$. The remaining two $\mathcal{N}=1$ chirals make a $\mathcal{N}=2$ hyper. Clearly, there are three ways to decompose the $\mathcal{N}=4$ multiplet in terms of $\mathcal{N}=2$ ones. Which decomposition is realized depends on the specific D-brane pair. Common transverse directions fit into the $\mathcal{N}=2$ vector, which features in the interaction.

Apart from the $N=4$ super Yang Mills fields, now there are additional fields coming from the strings stretched between two different stacks.
The massless modes of these strings lead to a $\mathcal{N}=2$ hypermultiplet in the bifundamental, containing two $\mathcal{N}=1$ chiral multiplets. For the  $(ij)^{th}$ pair of D-branes, we denote these  $\mathcal{N}=1$ chirals by $Z^{(ij)} = (z^{(ij)}, \chi^{(ij)}), Z^{(ji)}= (z^{(ji)}, \chi^{(ji)})$. They carry opposite charges.  

For  definiteness, let us consider the pair $(12)$, where the first stack of D2 branes wraps the $(45)$ cycle and second stack wraps the $(67)$ cycle. Then the relevant  $\mathcal{N}=2$ vector (of either branes) contains  $V=(A^\mu, \lambda)$ and $\Phi_3 = (\phi_3, \psi_3)$. We look at the supersymmetry from the perspective of the first brane.

Transverse separations appear through the supermultiplets $V^{(1)} - V^{(2)}$ and $\Phi^{(1)}_3 - \Phi^{(2)}_3$, where the superscript denotes the stack index.
The combination $V^{(1)} - V^{(2)}$ appears in the kinetic term, whereas the superpotential
\begin{align}
W_{(12)} &=  \sqrt{2}  \, Z^{(12)} (\Phi^{(2)}_3 - \Phi^{(1)}_3) Z^{(21)} \, .
\end{align}
fashions $\Phi^{(2)}_3 - \Phi^{(1)}_3$. The $\mathcal{N}=2$ hypermultiplet comprising $\Phi_1, \Phi_2$ does not appear in the interaction.

The $\mathcal{N}=2$ algebra in turn contains two $\mathcal{N}=1$ subalgebras, related through the $SU(2)_R$ R-symmetry, which in particular involves the exchange
\begin{align}
z^{(12)} \leftrightarrow -i (z^{(21)})^\dagger \, , \, \psi_3^{(k)} \leftrightarrow \lambda^{(k)} \, , \, \, k =1,2 \, . \label{su2R2stack}
\end{align}
In other  words, the Lagrangian remains invariant, if one uses the following $\mathcal{N}=1$ multiplets 
\begin{align}
\nn
\tilde{V}^{(k)} &= (A^{(k)}, \psi_3^{(k)}) , \, \tilde{\Phi}^{(k)}_3 = (\phi^{(k)}_3, \lambda^{(k)})  \, , \\
\nn
\tilde{\Phi}^{(k)}_1 &= (-i \left( \phi^{(k)}_2 \right) ^\dagger, \psi^{(k)}_1) \, , \, \tilde{\Phi}^{(k)}_2 = (-i \left( \phi^{(k)}_1 \right)^\dagger, \psi^{(k)}_2) \, , \, \, k=1,2, \, , \\
\tilde{Z}^{(12)} &= ( -i (z^{(21)})^\dagger, \chi^{(12)}), \, \tilde{Z}^{(21)} = ( -i (z^{(12)})^\dagger, \chi^{(21)}) \, . \label{flipfield}
\end{align}
$A$ denotes the gauge field.
%

For other pairs, the gauge interactions are straight forward, whereas the superpotentials respectively read
\begin{align}
\nn
W_{12} &=  \sqrt{2}  \, Z^{(12)} \left( \Phi^{(2)}_3 - \Phi^{(1)}_3 \right) Z^{(21)} \, , \\
\nn
W_{23} &= \sqrt{2} \, Z^{(32)} \left(  \Phi_1^{(2)} - \Phi_1^{(3)}\right) Z^{(23)}  \, , \\
\nn
W_{31} &= \sqrt{2}  \, Z^{(13)} \left(  \Phi_2^{(3)} -  \Phi_2^{(1)}  \right) Z^{(31)} \, , \\
\nn
W_{14} &= \sqrt{2}  \, Z^{(41)} \left(  \Phi_1^{(1)} -  \Phi_1^{(4)} \right) Z^{(14)}  \, , \\
\nn
W_{24} &= \sqrt{2}  \, Z^{(42)} \left(  \Phi_2^{(2)} - \Phi_2^{(4)}  \right) Z^{(24)}  \, , \\
W_{34} &= \sqrt{2}  \, Z^{(43)} \left(  \Phi_3^{(3)} - \Phi_3^{(4)}  \right) Z^{(34)}   \, .   \label{suppair}
\end{align}
\paragraph{Three stacks of D-branes}
Now let us add a third stack, say the D2 brane stack wrapping $x^8 \textendash x^9$.
 Three stacks altogether preserve $\mathcal{N}=1$ supersymmetry. There are two distinct ways this  could be done: either by using the multiplets $(V^{(k)}, \Phi_3^{(k)}, Z^{(12)}, Z^{(21)})$, or by using the multiplets $(\tilde{V}^{(k)}, \tilde{\Phi}^{(k)}_3, \tilde{Z}^{(12)}, \tilde{Z}^{(21)})$, where $k=1,2$. 
Reduced supersymmetry 
opens the doors to new terms in  the superpotentials $W_{lin}$, linear in $\Phi$, and $W_{cub}$, cubic in $Z$. The choice of the particular $\mathcal{N}=1$ subalgebra manifests in additional terms in the superpotential, where either $\Phi^{(k)}_3, Z^{(12)}, Z^{(21)}$ appears, or $\tilde{\Phi}^{(k)}_3, \tilde{Z}^{(12)}, \tilde{Z}^{(21)}$. Either way, the difference can be washed away by renaming fields.
Note, unlike in the case of higher supersymmetries, these terms are not fixed by supersymmetry. 
Reduced supersymmetry also allows for Fayet Iliopoulos (FI) parameter $\zeta^{(i)}$ for the $i^{th}$ stack \cite{Fayet:1974jb}.  

%
\paragraph{Four stacks of D-branes}
The orientation of the fourth brane is no longer inconsequential. Depending on whether the condition \refb{smatch} is satisfied or not, the whole system would preserve either $\mathcal{N}=1$ supersymmetry or will fail to preserve any supersymmetry. 

Preservation of $\mathcal{N}=1$ supersymmetry implies that the total Lagrangian would be expressible in terms of $\mathcal{N}=1$ superfields. This  can simply be achieved by using superfields $( V^{(k)} , \Phi_1^{(k)} , \Phi_2^{(k)} , \Phi_3^{(k)} )$ for the $k^{th}$ stack and the superfields $( Z^{(kl)}, Z^{(lk)})$ for the $(kl)^{th}$ pair of stacks. Apart from the pairwise pieces \refb{suppair}, the superpotential comprises the following pieces
 \begin{align}
\nn
W_{lin} &= \sqrt{2} \Big( c_{12} \left( \Phi_3^{(1)} -  \Phi_3^{(2)} \right) +
c_{23} \left(  \Phi_1^{(2)} -  \Phi_1^{(3)} \right) +
c_{31} \left(  \Phi_2^{(3)} - \Phi_2^{(1)} \right) \\
&+ c_{14} \left( \Phi_1^{(1)} - \Phi_1^{(4)} \right) +
c_{24} \left( \Phi_2^{(2)} - \Phi_2^{(4)} \right)  +
c_{34} \left( \Phi_3^{(3)} - \Phi_3^{(4)} \right)  \Big) \, ,  \label{wlin} \\
\nn
W_{cub} &= \sqrt{2} \Big[ 
Z^{(31)} Z^{(12)} Z^{(23)} + Z^{(13)} Z^{(32)} Z^{(21)} + Z^{(12)} Z^{(24)} Z^{(41)} + Z^{(42)} Z^{(21)} Z^{(14)} \\
& - Z^{(13)} Z^{(34)} Z^{(41)} + Z^{(31)} Z^{(14)} Z^{(43)} + Z^{(34)} Z^{(42)} Z^{(23)} + Z^{(43)} Z^{(32)} Z^{(24)}   
\Big] \label{cubiic}
\end{align} 
From the gauge interactions of the superfield $Z^{(kl)}$ will involve $V^{(k)} - V^{(l)}$.
This case has been shown to reproduce the correct degeneracies \cite{Chowdhury:2014yca, Chowdhury:2015gbk}.

\section{The 4-charge extremal non-BPS black hole} \label{snsusy}
We set out by noting that the worldline Lagrangian should satisfy the following properties:
\begin{enumerate}
\item
Over all there should be 28 Goldstones and 32 Goldstinos. \label{1}
\item
\begin{enumerate}
\item
each stack should preserve 16 supercharges, \label{2a}
\item
each pair of stacks should preserve 8 supercharges, \label{2b}
\item 
each triplet of stacks should preserve 4 supercharges, \label{2c}
\item 
four triplets of stacks should preserve four different supersymmetries.  \label{2d}
\end{enumerate}
\end{enumerate}
Number of Goldstones is the same as the 1/8 BPS brane system, as two systems differ only by change of orientation of some D-brane stacks. Number of Goldstinos reflect the fact that all 32 supersymmetries of type IIA string theory are broken. Having different number of Goldstones and Goldstinos is challenging since condition \refb{2a}, \refb{2b} entails that the Lagrangian is still written in terms of superfields. 
Satisfaction of condition \refb{2c} is straight forward, whereas condition \refb{2d} is non-trivial. To appreciate this, let us denote the $SU(2)_R$ symmetry corresponding to the pair $(kl)$ as $R_{kl}$. Let the (123) part of the Lagrangian fashion the fields $(V^{(k)}, \Phi^{(k)}), k=1,2,3$ and $Z^{(kl)}, k,l=1,2,3, k \neq l$. Then demanding that the triplet (124) preserves different supersymmetry than the triplet $(123)$ implies the pairs $(14)$ and $(24)$ should involve fields $R_{12} (V^{(k)}, \Phi_1^{(k)}, \Phi_2^{(k)}, \Phi_3^{(k)}), k=1,2$. In particular this means that $R_{12}  \Phi^{(1)}_1$ and $R_{12}  \Phi^{(2)}_2$ should feature in the superpotentials $\tilde{W}_{14}$ and $\tilde{W}_{24}$ respectively. Here  $\tilde{W}_{kl}$ stands for the analog of $W_{kl}$ for the present case. On the other hand, demanding (123) and (134) preserves different supersymmetries in particular entails that  $\tilde{W}_{14}$ should involve $R_{13} \Phi^{(1)}_1$. Since $R_{12}  \Phi^{(1)}_1 \neq R_{13} \Phi^{(1)}_1$, simultaneous satisfaction of \refb{2d} proves tricky. 

The solution is to augment the $SU(2)_R$ rotations by appropriate $SU(4)_R$ rotations for the adjoint fields. To this end 
%
%
let us first define the following multiplets
\begin{align}
\nn
V &= (A, \lambda), \, \Phi_1 = (\phi_1, \psi_1), \, \Phi_2 = (\phi_2, \psi_2), \,  \Phi_3 = (\phi_3, \psi_3), \, \\
\nn
\tilde{V} &= (A, \psi_1), \, \tilde{\Phi}_1 = (\phi_1, \lambda), \, \tilde{\Phi}_2 = (- i \phi_2^\dagger, \psi_3), \,  \tilde{\Phi}_3 = (- i \phi_3^\dagger, \psi_2), \, \\
\nn
\tilde{\tilde{V}} &= (A, \psi_2), \, \tilde{\tilde{\Phi}}_1 = (- i \phi_1^\dagger, \psi_3), \, \tilde{\tilde{\Phi}}_2 = ( \phi_2, \lambda), \,  \tilde{\tilde{\Phi}}_3 = (- i \phi_3^\dagger, \psi_1), \, \\
\tilde{\tilde{\tilde{V}}} &= (A, \psi_3), \, \tilde{\tilde{\tilde{\Phi}}}_1 = (- i \phi_1^\dagger, \psi_2), \, \tilde{\tilde{\tilde{\Phi}}}_2 = (- i \phi_2^\dagger, \psi_1), \,  \tilde{\tilde{\tilde{\Phi}}}_3 = ( \phi_3, \lambda). \, 
\end{align}
For the sake of brevity we shall specify the choice of these multiplets respectively as $(V, \Phi)$, $(\tilde{V}, \tilde{\Phi})$, $( \tilde{\tilde{V}}, \tilde{\tilde{\Phi}})$, $(\tilde{\tilde{\tilde{V}}}, \tilde{\tilde{\tilde{\Phi}}})$. Note that the $\mathcal{N}=4$ parts of the Lagrangian are agnostic to which of these supermultiplets are used.
Let us denote the R-symmetry that rotates the fermions of $V$ and $\Phi_i$ as $R_i$. Define $\tilde{R}_i, \tilde{\tilde{R}}_i, \tilde{ \tilde{ \tilde{R}}}_i$ similarly. Then it can be checked that up to renomenclature
\begin{align}
\nn
R_1 (V, \Phi) &= (\tilde{V}, \tilde{\Phi}) \, , \, R_2 (V, \Phi) = (\tilde{\tilde{V}}, \tilde{\tilde{\Phi}}) \, , \, R_3 (V, \Phi) = (\tilde{\tilde{\tilde{V}}}, \tilde{\tilde{\tilde{\Phi}}}) \, , \\
\nn
\tilde{R}_1 (\tilde{V}, \tilde{\Phi}) &= (V, \Phi) \, , \, \tilde{R}_2 (\tilde{V}, \tilde{\Phi}) =  (\tilde{\tilde{\tilde{V}}}, \tilde{\tilde{\tilde{\Phi}}}) \, , \, \tilde{R}_3 (\tilde{V}, \tilde{\Phi}) = (\tilde{\tilde{V}}, \tilde{\tilde{\Phi}}) \, , \\
\nn
\tilde{\tilde{R}}_1 (\tilde{\tilde{V}}, \tilde{\tilde{\Phi}}) &= (\tilde{\tilde{\tilde{V}}}, \tilde{\tilde{\tilde{\Phi}}}) \, , \, \tilde{\tilde{R}}_2 (\tilde{\tilde{V}}, \tilde{\tilde{\Phi}}) = (\tilde{V}, \tilde{\Phi}) \, , \, \tilde{\tilde{R}}_1 (\tilde{\tilde{V}}, \tilde{\tilde{\Phi}}) = (\tilde{V}, \tilde{\Phi}) \, , \\
\tilde{\tilde{\tilde{R}}}_1 (\tilde{\tilde{\tilde{V}}}, \tilde{\tilde{\tilde{\Phi}}}) &= (\tilde{\tilde{V}}, \tilde{\tilde{\Phi}}) \, , \, \tilde{\tilde{\tilde{R}}}_2 (\tilde{\tilde{\tilde{V}}}, \tilde{\tilde{\tilde{\Phi}}}) =  (\tilde{V}, \tilde{\Phi}) \, , \, \tilde{\tilde{\tilde{R}}}_3 (\tilde{\tilde{\tilde{V}}}, \tilde{\tilde{\tilde{\Phi}}}) = (V, \Phi) \, . \label{RRR}
\end{align}
Note $R_i, \tilde{R}_i, \tilde{\tilde{R}}_i, \tilde{\tilde{\tilde{R}}}_i$, are respectively associated with the brane pairs $(i,4)$ and $(j,k)$ where $i,j,k=1,2,3$ and $i \neq j \neq k$. 

The superpotential associated to various pairs are 
\begin{align}
\nn
\tilde{W}_{12} &= \sqrt{2} Z^{(12)} \left( \Phi_3^{(1)} - \Phi_3^{(2)} \right) Z^{(21)} \, , \\
\nn
\tilde{W}_{23} &= \sqrt{2} Z^{(23)} \left( \Phi_1^{(2)} - \Phi_1^{(3)} \right) Z^{(32)} \, , \\
\nn
\tilde{W}_{31} &= \sqrt{2} Z^{(31)} \left( \Phi_2^{(3)} - \Phi_2^{(1)} \right) Z^{(13)} \, , \\
\nn
\tilde{W}_{14} &= \sqrt{2} Z^{(14)} \left( \tilde{\tilde{\tilde{\Phi}}}_1^{(1)} - \tilde{\tilde{\tilde{\Phi}}}_1^{(4)} \right) Z^{(41)} \, , \\
\nn
\tilde{W}_{24} &= \sqrt{2} Z^{(24)} \left( \tilde{\tilde{\tilde{\Phi}}}_2^{(2)} - \tilde{\tilde{\tilde{\Phi}}}_2^{(4)} \right)  Z^{(42)} \, , \\
\tilde{W}_{34} &= \sqrt{2} Z^{(34)} \left( \tilde{\Phi}_3^{(3)} - \tilde{\Phi}_3^{(4)} \right) Z^{(43)} \, .   \label{suppairnsusy}
\end{align}
In the gauge interaction of these pairs, we respectively have $V^{(1)} - V^{(2)}, \, V^{(2)} - V^{(3)}, \, V^{(3)} - V^{(1)}, \, \tilde{\tilde{\tilde{V}}}^{(1)} - \tilde{\tilde{\tilde{V}}}^{(4)}, \tilde{\tilde{\tilde{V}}}^{(2)} - \tilde{\tilde{\tilde{V}}}^{(4)}, \, \tilde{V}^{(3)} - \tilde{V}^{(4)}$.

The Goldstones are same as the supersymmetric case
\begin{align}
\vec{X}^{(1)} + \vec{X}^{(2)} + \vec{X}^{(3)} + \vec{X}^{(4)}, \, \phi_3^{(1)} + \phi_3^{(2)} , \, \phi_1^{(2)} + \phi_1^{(3)}, \,  \phi_2^{(3)} + \phi_2^{(1)} , \, \phi_1^{(1)} + \phi_1^{(4)}, \, \phi_2^{(2)} + \phi_2^{(4)}, \,  \phi_3^{(3)} + \phi_3^{(4)} \, . 
\end{align}
Note the vector multiplet actually leads to two Goldstones, not three, since the Gauss law constraint negates one.

Coming to Goldstinos we first note that only the following  combinations of fermions appear in interaction
\begin{align}
\nn
&\psi_3^{(1)} - \psi_3^{(2)}, \, \psi_1^{(2)} - \psi_1^{(3)}, \,  \psi_2^{(3)} - \psi_2^{(1)} \, , \\ 
\nn
&\psi_2^{(1)} - \psi_2^{(4)}, \, \psi_1^{(2)} - \psi_1^{(4)}, \,  \psi_2^{(3)} - \psi_2^{(4)} \, , \\
\nn
&\psi_3^{(1)} - \psi_3^{(4)}, \, \psi_3^{(2)} - \psi_3^{(4)}, \,  \psi_1^{(3)} - \psi_1^{(4)} \, , \\
&\lambda^{(1)} - \lambda^{(2)}, \, \lambda^{(2)} - \lambda^{(3)}, \, \lambda^{(3)} - \lambda^{(1)} \, ,
\end{align}
implying that the remaining combinations are Goldstinos
\begin{align}
\nn
\psi^{(1)}_1 \, , \psi^{(2)}_1 + \psi^{(3)}_1 + \psi^{(4)}_1 \, , \\
\nn
\psi^{(2)}_2 \, , \psi^{(1)}_2 + \psi^{(3)}_2 + \psi^{(4)}_2 \, , \\
\nn
\psi^{(3)}_3 \, , \psi^{(1)}_3 + \psi^{(2)}_3 + \psi^{(4)}_3 \, , \\
\lambda^{(4)} \, , \lambda^{(1)} + \lambda^{(2)} + \lambda^{(3)} \, .
\end{align}
We see there are $32$ Goldstinos, signifying spontaneous breaking of all 32 supersymmetries.

At this stage, let us check that various supersymmetry requirements are met. Pairwise supersymmetry is there by construction. Coming to triplets, the supersymmetry of $(123)$ pair is apparent. 
Supersymmetry of (124) is not apparent since $\tilde{W}_{12}$ fashions $(V, \Phi)$ superfields, whereas $\tilde{W}_{14}, \tilde{W}_{24}$ fashion $\left( \tilde{\tilde{\tilde{V}}}, \tilde{\tilde{\tilde{\Phi}}} \right)$ superfields. Acting $R_3$ on $\tilde{W}_{12}$ expresses all of them in terms of $\left( \tilde{\tilde{\tilde{V}}}, \tilde{\tilde{\tilde{\Phi}}} \right)$ superfields, exhibiting that the triplet (124) preserves $\mathcal{N}=1$ supersymmetry, but a different one compared to the triplet (123).  Note $R_3$ also acts on $Z^{12}, Z^{21}$. 


To see the supersymmetry of (134), one has to $SU(2)_R$ rotate $\tilde{W}_{13}, \tilde{W}_{14}, \tilde{W}_{34}$ respectively by $R_2, \tilde{\tilde{\tilde{R}}}_1, \tilde{R}_3$. These are precisely the $SU(2)_R$ rotations associated with those brane pairs. Similarly to see the supersymmetry of the triplet (234), one has to $SU(2)_R$ rotate $\tilde{W}_{23}, \tilde{W}_{24}$ respectively by $R_1, \tilde{\tilde{\tilde{R}}}_2$. The relative $SU(2)_R$ rotation $\tilde{R}_3$ between (134) and (234) fields imply these  triplets preserve different $\mathcal{N}=1$ supersymmetries.
To summarize, the triplets (123), (234), (134), (124),  make supersymmetry apparent respectively in terms of $(V, \Phi), \,  \left( \tilde{V}, \tilde{\Phi} \right), \,  \left( \tilde{\tilde{V}}, \tilde{\tilde{\Phi}} \right),$ and $\left( \tilde{\tilde{\tilde{V}}}, \tilde{\tilde{\tilde{\Phi}}} \right)$.

For the analog of the cubic superpotential  \refb{cubiic}, we note each piece corresponds to a triplet. Supersymmetry considerations of various 
triplets lead to the following cubic superpotential 
\begin{align}
\nn
\tilde{W}_{cub} &= \sqrt{2} \Big[ 
Z^{(31)} Z^{(12)} Z^{(23)} + Z^{(13)} Z^{(32)} Z^{(21)} + \tilde{Z}^{(12)} Z^{(24)} Z^{(41)} + Z^{(42)} \tilde{Z}^{(21)} Z^{(14)} \\
& - \tilde{Z}^{(13)} \tilde{Z}^{(34)} \tilde{Z}^{(41)} + \tilde{Z}^{(31)} \tilde{Z}^{(14)} \tilde{Z}^{(43)} + Z^{(34)} \tilde{Z}^{(42)} \tilde{Z}^{(23)} + Z^{(43)} \tilde{Z}^{(32)} \tilde{Z}^{(24)}   
\Big] \, , \label{tcubic}
\end{align}
where $\tilde{Z}^{(ij)}$ denotes the appropriate $SU(2)_R$ rotated form of $Z^{(ij)}$. Clearly, the triplets $(ijk)$ and $(ijl)$, $i,j,k,l=1,2,3,4,\, i \neq j \neq k \neq l$ preserve different supersymmetries.

Finding counterpart of the linear superpotential $W_{lin}$ \refb{wlin} proves tricky through. E.g. consider the piece $ \sqrt{2} c_{12} \left( \Phi_3^{(1)} - \Phi_3^{(2)} \right)$ associated with the pair (12). This piece explicitly preserves a particular $\mathcal{N}=1$ subgroup of $\mathcal{N}=2$ supersymmetry preserved by the pair (12). In the non-supersymmetric case, this subgroup is preserved either by the triplet (123) or by (124) but not both. Therefore this piece is bound to be inconsistent with the supersymmetry of either (123) or (124). Similar problem exists for other pieces of \refb{wlin}. It is not clear which superfields should appear in the superpotential. We postulate the most general linear superpotential consistent with the Goldstones and Goldstinos
\begin{align}
\nn
\tilde{W}_{lin} &= \sqrt{2} \Big( c'_{12} \left( \Phi_3^{(1)} - \Phi_3^{(2)} \right) +
c'_{23} \left( \Phi_1^{(2)} - \Phi_1^{(3)} \right) +
c'_{31} \left( \Phi_2^{(3)} - \Phi_2^{(1)} \right) \\
\nn
&+ c'_{14} \left( \tilde{\tilde{\tilde{\Phi}}}_1^{(1)} - \tilde{\tilde{\tilde{\Phi}}}_1^{(4)} \right) +
c'_{24} \left( \tilde{\tilde{\tilde{\Phi}}}_2^{(2)} - \tilde{\tilde{\tilde{\Phi}}}_2^{(4)} \right)  +
c'_{34} \left( \tilde{\Phi}_3^{(3)} - \tilde{\Phi}_3^{(4)} \right)  \, \\ 
\nn
&+  c''_{12} \left( \tilde{\tilde{\tilde{\Phi}}}_3^{(1)} - \tilde{\tilde{\tilde{\Phi}}}_3^{(2)} \right) +
c''_{23} \left( \tilde{\Phi}_1^{(2)} - \tilde{\Phi}_1^{(3)} \right) +
c''_{31} \left( \tilde{\tilde{\Phi}}_2^{(3)} - \tilde{\tilde{\Phi}}_2^{(1)} \right) \\
&+ c''_{14} \left( \tilde{\tilde{\Phi}}_1^{(1)} - \tilde{\tilde{\Phi}}_1^{(4)} \right) +
c''_{24} \left( \tilde{\Phi}_2^{(2)} - \tilde{\Phi}_2^{(4)} \right)  +
c''_{34} \left( \tilde{\tilde{\Phi}}_3^{(3)} - \tilde{\tilde{\Phi}}_3^{(4)} \right)  \Big) \,  . \label{twlin}
\end{align}

\paragraph{The potential:} The potential has the form 
\begin{align}
V &= V_{gauge} + V_D + V_F \, .
\end{align}
Here $V_{gauge}$ comes from the spatial part of the gauge kinetic term and is a sum of pairwise terms $|\vec{X}^{(i)} - \vec{X}^{(j)} |^2 |Z^{(ij)}|^2$. For generic values of $Z$ fields, this can be set to zero by taking $\vec{X}^{(i)} - \vec{X}^{(j)} = \vec{0} \, \forall i \neq j$, i.e. all branes to coincide along the flat directions. Minimization of in F and D term potentials are not affected by this, since $\vec{X}$ fields do not appear there. 

 The D-term potential $V_D$ is entirely fixed by the interaction of the gauge multiplets and hence is same as in the BPS case. The FI parameters were determined solely in terms of the background fields in the BPS case \cite{Chowdhury:2014yca, Chowdhury:2015gbk}. We expect this to continue in the present case.

Finally we come to the F-term potential $V_F$, the detail of which are relegated to Appendix \ref{appF}. Most important feature of F-term equations is that they are no more holomorphic. This immediately requires a change of strategy in finding minima as compared to the supersymmetric case. In the supersymmetric case, the F-term equations are invariant under not only the $U(1) \times U(1) \times U(1)$ gauge symmetry, but an enhanced $\mathbb{C}^\star \times \mathbb{C}^\star \times \mathbb{C}^\star$ symmetry. The effect of D-term equations can be incorporated as gauging this enhanced symmetry. Thus it suffices to solve the F-term equations alone subject to the complexified gauge  invariance. But now due to lack of holomorphicity there is no such enhanced symmetry. So one must write down the whole potential and find its minima numerically. 

\paragraph{The ground state:} It is not clear if the classical minima are degenerate. If not, quite clearly one has a unique ground state. Even if they are degenerate they may or may not have vanishing energy. If the global minima itself has a non-zero energy, that would imply that microscopically there are no extremal configurations even at classical level. Although we feel this is unlikely, our preliminary numerical search using Python did not find any zero energy minima. Most interesting possibility is that of large number of degenerate minima at zero energy. Now that there is no supersymmetry, quantum mechanically this degeneracy will be lifted leading to a unique ground state. Moreover, as shown in \cite{Mondal:2023bcx} this would lead to exponentially many low lying states, separated from the rest of the spectrum by a gap. Such spectra in particular leads to $S \sim \log{} T$ in a low but not too low temperature regime.


\section{Discussion} \label{sdisc}
Besides presenting concrete evidence in the microscopic side that extremal non-supersymmetric black holes indeed have a unique ground state, our work also offers an explicit microscopic set up for studying statistical mechanics of black holes. Many questions are raised too. 

The non-supersymmetric nature of the system entails that individual theories on various brane worldvolumes will undergo non-trivial renomalization group flow, which in turn will modify the quantum mechanics capturing the zero mode sector. Whereas it would be interesting to compute these modifications, we do not expect these to alter the essential physics. 

Coming to the elephant in the room, namely the extremal entropy, we note now that supersymmetry is gone, there are no protected quantities. Nevertheless given numerous instances of  microstate counting accounting for non-supersymmetric entropy one would be hopeful that some notion of extremal entropy should persist in the microscopic description. Furthermore it should be fairly close to the Bekenstein-Hawking entropy for large black holes. However the sheer absence of any truly extremal states raises the question- how should one define the extremal entropy in the microscopic description? Should the potential minima have zero energy, then the logarithm of the number of minima would be a natural candidate for the extremal entropy at classical level, whereas at the quantum level the scenario proposed in  \cite{Mondal:2023bcx} will prevade. On the other hand if a significant fraction of the classical minima fail to attain zero energy, the situation is far from clear. 


The angular momentum of the microstates is also of interest. For supersymmetric black holes this can be defined using Lefschetz SU(2) \cite{Denef:2002ru, Bena:2012hf} and leads to the zero angular momentum conjecture \cite{Denef:2002ru, Sen:2009vz, Dabholkar:2010rm, Sen:2011ktd, Bringmann:2012zr, Chowdhury:2015gbk} . But the extension of this notion to the present case is not straight forward, as the vacuum manifold is not a complex manifold. 
 
Lastly, two extremal black holes differing only in the signs of the charges have the same surface area hence same Bekenstein Hawking entropies. Although quantum entropy functions would be different, for large charges the difference should be subleading. This implies that the extremal non-supersymmetric entropy should not be too far from the that of the corresponding supersymmetric black hole. This may be useful in coming up with the definition of extremal non-supersymmetric entropy in the present microscopic description.

\paragraph{Acknowledgement:}  This work was supported by Banaras Hindu University. It is a pleasure to thank all the participants of the Solvay workshop on near-extremal black holes at Brussels, 2024 for useful discussions. 
I am indebted to Anirban Basu, Anshuman Maharana,  Ashoke Sen, Dileep Jatkar, Roberto Emparan and Simon Jones for valuable comments and discussions. 
\newpage

\appendix
\section{F-term equations} \label{appF}
Using the fact that a chiral superfield and its $SU(2)_R$ rotated form has same auxiliary field, we get the following F-term equations:
\begin{align}
z^{(ij)} z^{(ji)} + c'_{ij} + c''_{ij} &= 0 \, , \quad \forall (ij) \label{f1}\\
\nn
 \left( \phi_3^{(1)} - \phi_3^{(2)} \right) z^{(21)} + z^{(23)} z^{(31)} + z^{(24)} z^{(41)} &= 0 \, , \\
 \nn
z^{(12)}  \left( \phi_3^{(1)} - \phi_3^{(2)} \right) + z^{(13)} z^{(32)} + z^{(14)} z^{(42)} &= 0 \, , \\
\nn
 \left( \phi_1^{(2)} - \phi_1^{(3)} \right) z^{(32)} +  z^{(31)} z^{(12)} -i z^{(34)} ( z^{(24)} )^\dagger &= 0 \, , \\
 \nn
z^{(23)} \left( \phi_1^{(2)} - \phi_1^{(3)} \right)  + z^{(21)} z^{(13)} - i ( z^{(42)} )^\dagger z^{(43)} &= 0 \, , \\
\nn
 \left( \phi_2^{(3)} - \phi_2^{(1)} \right) z^{(13)} + z^{(12)} z^{(23)}  - ( z^{(41)} )^\dagger ( z^{(34)} )^\dagger &= 0 \, , \\
 \nn
z^{(31)} \left( \phi_2^{(3)} - \phi_2^{(1)} \right) + z^{32} z^{21} + ( z^{43} )^\dagger ( z^{14} )^\dagger  &= 0 \, , \\
\nn
- i \left( \phi_1^{(1)} - \phi_1^{(4)} \right)^\dagger z^{(41)} - i z^{42} ( z^{12} )^\dagger - ( z^{34})^\dagger ( z^{13} )^\dagger &= 0 \, , \\
\nn
- i z^{(14)} \left( \phi_1^{(1)} - \phi_1^{(4)} \right)^\dagger -i ( z^{21} )^\dagger z^{24} + ( z^{31})^\dagger ( z^{43} )^\dagger  &= 0 \, , \\
\nn
 \left(  \phi_2^{(2)} - \phi_2^{(4)}  \right) z^{(24)} - i ( z^{12} )^\dagger z^{14} - i ( z^{32} )^\dagger z^{34}  &= 0  \, , \\
 \nn
z^{(42)} \left(  \phi_2^{(2)} - \phi_2^{(4)}  \right) - i z^{41} ( z^{21} )^\dagger - i z^{43} ( z^{23} )^\dagger  &= 0 \, , \\
\nn
- i \left( \phi_3^{(3)} - \phi_3^{(4)} \right)^\dagger z^{(43)} + ( z^{14} )^\dagger ( z^{31} )^\dagger - ( z^{24} )^\dagger ( z^{32} )^\dagger  &= 0 \, , \\
- i z^{(34)} \left( \phi_3^{(3)} - \phi_3^{(4)} \right)^\dagger - ( z^{13} )^\dagger ( z^{41} )^\dagger  - ( z^{23} )^\dagger ( z^{42} )^\dagger &= 0 \, . \label{f2}
\end{align}
Note, \refb{f1} stands for 6 equations, whereas \refb{f2} equations fix the $\phi$ fields in terms of $z$ fields. Each $\phi$ is actually fixed
by two equations. So we get a bunch of consistency conditions on $z$ fields, which could be treated as effective F-term equations.

\bibliography{nExt} 
\bibliographystyle{JHEP}   

\end{document}